\documentclass[twocolumn,preprintnumbers,aps]{revtex4}
\usepackage{graphicx}
\begin{document}

\preprint{ For Physical Review B }
\title{ Multiband  theory of multi-exciton complexes in self-assembled quantum dots }

\author{ Weidong Sheng }
\email{ weidong.sheng@nrc-cnrc.gc.ca }
\author{ Shun-Jen Cheng }
\altaffiliation[Current address: ]{Electrophysics Department, National Chiao Tung University, Hsinchu 30050, Taiwan,
Republic of China.}
\author{ Pawel Hawrylak }

\affiliation{ Institute for Microstructural Sciences, National Research Council of Canada, Ottawa, ON K1A 0R6, Canada }

\begin{abstract}
We report on a multiband microscopic theory of many-exciton complexes in self-assembled quantum dots. The single
particle states are obtained by three methods: single-band effective-mass approximation, the multiband $k\cdot p$
method, and the tight-binding method. The electronic structure calculations are coupled with strain calculations via
Bir-Pikus Hamiltonian. The many-body wave functions of $N$ electrons and $N$ valence holes are expanded in the basis of
Slater determinants. The Coulomb matrix elements are evaluated using statically screened interaction for the three
different sets of single particle states and the correlated $N$-exciton states are obtained by the configuration
interaction method. The theory is applied to the excitonic recombination spectrum in InAs/GaAs self-assembled quantum
dots. The results of the single-band effective-mass approximation are successfully compared with those obtained by
using the of $k\cdot p$ and tight-binding methods. 
\end{abstract}
\maketitle

\section{Introduction}

Semiconductor self-assembled quantum dots (SAQDs)\cite{QD:Jacak, QD:Bimberg,QD:Hawrylak} are islands of one
semiconductor, e.g. InAs, in a host matrix of another semiconductor, e.g. GaAs. The elementary excitations, electrons
and holes, are believed to be confined in all three dimensions by the band gap difference between island and matrix
materials. The picture of electrons and holes as confined elementary excitations with effective mass, interacting via
Coulomb interactions has been successfully applied toward the explanation of many experimental results
\cite{QD:exciton, EXP:Drexler, EXP:Skolnick, EXP:Cingolani, EXP:Bayer2000, MultiCI:Hawrylak2000, MultiCI:Hartmann,
EXP:Karrai}. It is important to establish to what extent the effective-mass picture is applicable to the description of
electronic states of self-assembled quantum dots by a systematic comparison of different approaches. The self-assembled
quantum dots plus the surrounding barrier material contain millions of atoms and the density functional {\it ab initio}
calculations are not possible yet. Hence in this work we compare two simplified approaches, the multiband $k\cdot p$
method and the tight-binding method with the predictions of the effective-mass calculations. The multiband $k\cdot p$
method \cite{KP:Bahder, KP:Jiang, KP:Pryor, KP:Sheng2002, KP:Stier, KP:Holma, Strain:Tadic} accounts for the proper
structure of the valence band, including heavy, light and spin split-off hole bands. It is however limited to the top
of the valence band, does not account for the atomistic character of the interfaces between the dot and barrier
material, and is expected to break down as the size of the nanostructure decreases. The atomistic structure of the
nanostructure is captured in either the tight-binding \cite{TB:Bryant,TB:Klimeck} or pseudopotential approaches as
developed by Zunger and co-workers \cite{Pseudo:Williamson}. The tight-binding approach chosen here is the effective
bond orbital model (EBOM) \cite{EBOM:Chang,EBOM:Nair,EBOM:Sun}, a version of $sp^3$ tight-binding models. The advantage
of EBOM is that it extrapolates to the $k\cdot p$ approach making a direct comparison possible. The disadvantage of
EBOM is that it misses the lack of inversion symmetry of zincblende structures.
		
The single particle energy levels are not measured directly. What is measured in, {\it e.g.} optical experiments, is
the emission from self-assembled quantum dots as a function of the excitation power, or the number of electrons and
holes in the dot. The electrons and holes interact and form multi-exciton complexes. Emission from multi-exciton
complexes has been measured by a number of groups \cite{EXP:Landin, EXP:Heitz, EXP:Cingolani, EXP:Bayer2000,
MultiCI:Hartmann, EXP:Raymond1997, EXP:Zrenner, EXP:Skolnick, EXP:Baier, EXP:Dalacu}. The higher the pumping intensity
is, the more excitons are involved, thus the emission from higher excited electron and hole states can be observed. The
multi-exciton emission spectra have been interpreted using quantum mechanical methods such as the Hartree-Fock method
and the configuration interaction method (CI) \cite{CI:Szabo}, in which the multi-exciton complex states are
constructed from single-particle states of the system. It is a challenge to combine realistic single-particle states
calculated for the million-atom structures with these quantum mechanical methods. A number of theoretical approaches
have been proposed to address this issue, such as combining multiband ${\bf k}\cdot{\bf p}$ single-particle states with
self-consistent Hartree-Fock method \cite{KP:Stier} and combining single-band effective-mass \cite{MultiCI:Barenco1995,
MultiCI:Wojs1996, MultiCI:Hawrylak1999, MultiCI:Brasken, MultiCI:Hartmann} or microscopic pseudopotential wave
functions \cite{MultiCI:Franceschetti, MultiCI:Bester}.

In this paper, we use a general approach which combines different multiband calculations of single-particle states
with the CI method for the calculation of multi-exciton states. By using single-particle states obtained
from the single-band effective-mass approximation, the multiband ${\bf k}\cdot{\bf p}$ method and the atomistic
tight-binding-like method, we are able to compare the multi-exciton emission spectra obtained from different
single-particle states and determine both the validity of the effective-mass approximation as well as
the validity of multi-exciton emission spectra as fingerprints of electronic structure of quantum dots.


\section{Single-particle calculation}

The single particle calculations for self-assembled quantum dots started with the effective-mass calculations which
related shape and size of the dots to the single particle energy levels \cite{EA:Wojs}. As experimental information
accumulated, more sophisticated approaches were developed, such as single-band effective-mass method coupled with
strain calculation \cite{EA:Fonseca, EA:Korkusinski,CI:Cheng}, eight-band ${\bf k}\cdot{\bf p}$ method \cite{KP:Pryor,
KP:Sheng2002, KP:Stier, KP:Holma}, tight-binding methods \cite{EBOM:Sun, TB:Klimeck} and the empirical pseudopotential
method \cite{Pseudo:Williamson}. In the following, we briefly describe the single-band effective-mass method, the
eight-band ${\bf k}\cdot{\bf p}$ method and EBOM for the calculation of single-particle states in SAQDs.


\subsection{Effective-mass single-particle states}

Here we use a single-band model with anisotropic effective masses of electrons and holes treated as
adjustable parameters. The Hamiltonians, including strain,
read
\begin{eqnarray}
\hat{H}_e &=& -\frac{\hbar^2}{2m^e_\parallel} ( \frac{\partial^2}{\partial x^2} + \frac{\partial^2}{\partial y^2} ) 
-\frac{\hbar^2}{2m^e_\perp} \frac{\partial^2}{\partial z^2} + a_c H_s + V_{bo}^e, \cr
\hat{H}_h &=& \frac{\hbar^2}{2m^h_\parallel} ( \frac{\partial^2}{\partial x^2} + \frac{\partial^2}{\partial y^2} ) 
+\frac{\hbar^2}{2m^h_\perp} \frac{\partial^2}{\partial z^2} \cr
&-& a_v H_s - bB_s + V_{bo}^h,
\label{EMA-Hamiltonian}
\end{eqnarray}
where $H_s = \varepsilon_{xx} + \varepsilon_{yy} + \varepsilon_{zz}$ and $B_s = \varepsilon_{zz} - \frac{1}{2}
(\varepsilon_{xx} + \varepsilon_{yy})$ is the hydrostatic and biaxial strain component, respectively, $V_{bo}^e$ and
$V_{bo}^h$ are the potentials from the band offsets between island (InAs) and matrix (GaAs) material. $a_c$, $a_v$, and
$b$ are the deformation potential parameters that are also used in the multiband ${\bf k}\cdot{\bf p}$ method and EBOM.


\subsection{Eight-band ${\bf k}\cdot{\bf p}$ single-particle states}

The eight-band ${\bf k}\cdot{\bf p}$ method uses eight Bloch functions at the $\Gamma$ point of the Brillouin zone as
basis functions to describe electron states with finite wavevector. As the lateral size of SAQDs is usually much
larger than the lattice constant, it has been widely used in the calculation of confined electron states in SAQDs
\cite{KP:Jiang, KP:Pryor, KP:Sheng2002}. In general, the multiband ${\bf k}\cdot{\bf p}$ Hamiltonian can be
written as
\begin{eqnarray}
\hat{H}_{{\bf k}\cdot{\bf p}} &=& {\bf E}_{bo} + {\bf A}_{x}\hat{k}_x\hat{k}_x + 
{\bf A}_{y}\hat{k}_y\hat{k}_y + {\bf A}_{z}\hat{k}_z\hat{k}_z \cr
&+& {\bf B}_{xy}\hat{k}_x\hat{k}_y + {\bf B}_{yz}\hat{k}_y\hat{k}_z + {\bf B}_{xz}\hat{k}_x\hat{k}_z \cr
&+& {\bf C}_x\hat{k}_x + {\bf C}_y\hat{k}_y + {\bf C}_z\hat{k}_z ,
\label{KP-Hamiltonian}
\end{eqnarray}
where ${\bf E}_{bo}$ is the matrix for the band offsets, ${\bf A}$'s, ${\bf B}$'s, and ${\bf C}$'s are the coefficient
matrices \cite{KP:Bahder}. By using the deformation potential theory, an additional part $\hat{H}_s$ \cite{KP:Bahder},
which has a similar form as $\hat{H}_{{\bf k}\cdot{\bf p}}$, can be added to take into account the effects of the
strain.

Fig. \ref{bands-diagram} plots the energy dispersion of GaAs bands (broken lines) calculated by the eight-band ${\bf
k}\cdot{\bf p}$ method. Note a spurious crossing between valence bands at wavevectors k halfway between the $\Gamma$
and $X$ points. This crossing may result in spurious valence band states in SAQDs. The problem can be artificially
removed by adding additional terms proportional to $k^4$ into the eight-band ${\bf k}\cdot{\bf p}$ Hamiltonian
\cite{KP:Holma}.


\subsection{Tight-binding single-particle states}

By using the same number of basis functions as the eight-band ${\bf k}\cdot{\bf p}$ method, EBOM is a $sp^3$
tight-binding method based on an effective $fcc$ lattice \cite{EBOM:Chang}, i.e., a pair of cation and anion in a
zinc-blende lattice is treated as a single super-atom. The Hamiltonian is given by
\begin{eqnarray}
&& \langle {\bf R}\alpha | \hat{H}_{EB} | {\bf R}^\prime \alpha^\prime \rangle =
E_{p} \delta_{{\bf R}{\bf R}^\prime} \delta_{\alpha\alpha^\prime} + \sum_\tau 
\delta_{{\bf R}-{\bf R}^\prime, \tau} \cdot \cr
&& \{ E_{xy} \tau_\alpha \tau_{\alpha^\prime} (1 - \delta_{\alpha\alpha^\prime}) +
[ E_{xx} \tau_\alpha^2 + E_{zz} (1 - \tau_\alpha^2) ] \delta_{\alpha\alpha^\prime} \} , \cr
&& \langle {\bf R}s | \hat{H}_{EB} | {\bf R}^\prime s^\prime \rangle = E_s \delta_{{\bf R}{\bf R}^\prime} + 
\sum_\tau E_{ss} \delta_{{\bf R}-{\bf R}^\prime, \tau} , \cr
&& \langle {\bf R}s | \hat{H}_{EB} | {\bf R}^\prime \alpha \rangle = \sum_\tau E_{sp} 
\tau_\alpha \delta_{{\bf R}-{\bf R}^\prime, \tau} ,
\label{EBOM-Hamiltonian}
\end{eqnarray}
where $| {\bf R} \alpha \rangle$ denotes an orbital $\alpha$ located at site ${\bf R}$. $E_s$, $E_p$, $E_{ss}$,
$E_{sp}$, $E_{xy}$, $E_{xx}$, and $E_{zz}$ are parameters that are chosen to reproduce the conduction-band effective
mass, band gap, spin-split energy, and Luttinger parameters.

\begin{figure}
\vspace{5mm}
  \includegraphics[width=3in]{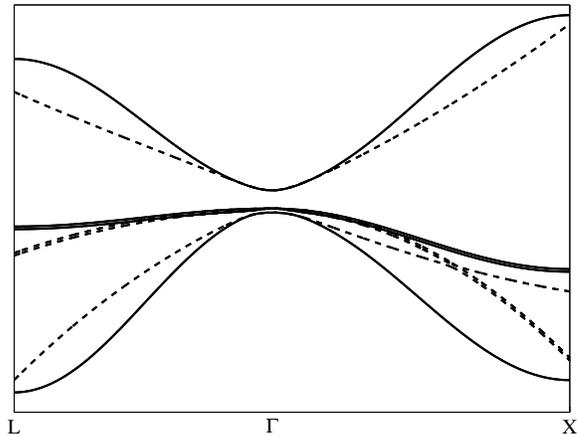}\\
  \caption{ Band structure of GaAs described by the tight-binding-like effective bond-orbital method (solid lines) and
the eight-band ${\bf k}\cdot{\bf p}$ method (dash lines).}
\label{bands-diagram}
\end{figure}

Fig. \ref{bands-diagram} plots the energy dispersion of GaAs bands (solid lines) calculated by EBOM. As EBOM uses the
same Luttinger parameters as the ${\bf k}\cdot{\bf p}$ theory does, both approaches give the same band structure near
the $\Gamma$ point. However, the spurious crossing from the eight-band ${\bf k}\cdot{\bf p}$ theory is not found in the
EBOM's band structure. Here, we adopted the two-center approximation in the parametrization \cite{EBOM:Nair} instead of
fitting the energy separation between the heavy-hole and the light-hole band at X point \cite{EBOM:Chang}.

Compared with the single-band calculation, the multiband methods gives more realistic confined states in SAQDs. These
states lack the symmetries, such as angular momentum and spin, which are usually preserved in the single-band
calculation. These symmetries are important in the CI calculation because they substantially reduce the total number of
configurations. In general, the multiband eigenstates of SAQDs do not have any spatial symmetry due to the effects of
shear strain even for a disklike or lens-shape dots that have circular symmetry. In addition, the total spin $S$ and
its projection $S_z$ are not conserved due to the spin-orbit interaction. However, we will show that the multiband
single-particle states are polarized and the polarization can be used to define quasi-spin.

\section{Formulation of the multi-exciton problem}

Our structure contains millions of atoms and hence millions of electrons. As long as the total system contains an
energy gap and a well defined ground state wavefunction, the intractable million electron problem can be replaced by a
much smaller problem of pairs of excitations in the form of quasi-electrons and quasi-holes. Formally, any electron
state can be expanded in terms of increasing number of pairs of excitations,
\begin{equation}
\Psi = \Psi_0 + \sum_{i,m} c_i^m \Psi_i^m + \sum_{ij,mn} c_{ij}^{mn} \Psi_{ij}^{mn} + \cdots ,
\label{excitation-expansion}
\end{equation}
where $\Psi_0$ is the Hartree-Fock ground state with all valence states occupied and conduction band empty. $\Psi_i^m$
is an excited state formed by removing an electron from the state $i$ in the valence band and creating a "hole", and
moving it to the state $m$ in the conduction band, creating an "electron". $\Psi_{ij}^{mn}$ is a doubly excited state
containing two "electrons" and two "holes", and so on. The number of electron-hole pairs is in principle not conserved
and this expansion can be used to describe all excited states \cite{MultiCI:Hawrylak1996}. However, in semiconductors,
the difference between the kinetic energies of different numbers of pair excitations is proportional to the band gap,
which is much larger than the Coulomb interaction mixing them. Therefore, different numbers of pair excitations are
practically independent from each other \cite{EX:Knox, EX:Haken, CI:Sherrill, MultiCI:Williamson}.
 
After solving the one electron problem (Eqs. [\ref{EMA-Hamiltonian}-\ref{EBOM-Hamiltonian}]) and obtaining the
single-particle eigenstates $\phi_i$ and their energies $E_i$, the Hamiltonian for the interacting electrons can be
written in second quantization as

\begin{equation}
\hat{H} = \sum_i E_i c_i^+ c_i + \frac{1}{2} \sum_{ijkl} V_{ijkl} c_i^+ c_j^+ c_k c_l.
\label{Hamiltonian}
\end{equation}
Here $V_{ijkl}$'s are the Coulomb matrix elements,
\begin{eqnarray}
V_{ijkl} &=& \int \int \phi_i^\ast({\bf r}_1) \phi_j^\ast({\bf r}_2) \frac{e^2}{4 \pi \epsilon({\bf
r}_1,{\bf r}_2) \cdot|{\bf r}_1-{\bf r}_2|} \cdot \cr
&&\phi_k({\bf r}_2) \phi_l({\bf r}_1) d{\bf r}_1 d{\bf r}_2 ,
\label{two-electron-integral}
\end{eqnarray}
$\epsilon({\bf r}_1,{\bf r}_2)$ is the dielectric function \cite{MultiCI:Franceschetti}. We replace it with the
dielectric constant $\epsilon$ throughout the calculation. The method for computation of these elements is given in the
appendix. The Hamiltonian for many-exciton complex can be written as \cite{MultiCI:Barenco1995, CI:Cheng},
\begin{eqnarray}
H_{ex} &=& \sum_i E_i^e c_i^+ c_i - \sum_i E_i^h h_i^+ h_i -
           \sum_{ijkl} V^{he}_{ijkl} h_i^{+} c_j^{+} c_k h_l \cr
       &+& \sum_{ijkl} X_{ijkl} h_i^{+} c_j^{+} c_k h_l + \frac{1}{2} \sum_{ijkl} V^{ee}_{ijkl} c_i^+ c_j^+ c_k c_l \cr
       &+& \frac{1}{2} \sum_{ijkl} V^{hh}_{ijkl} h_i^+ h_j^+ h_k h_l .
\label{second-quantization} 
\end{eqnarray}

The electron-hole exchange interaction elements, $X_{ijkl}$ are defined by
\begin{eqnarray}
X_{ijkl} &=& \int \int \phi_i^\ast({\bf r}_1) \phi_j^\ast({\bf r}_2)
\frac{e^2}{4 \pi \epsilon({\bf r}_1,{\bf r}_2) \cdot|{\bf r}_1-{\bf r}_2|} \cdot \cr
&& \phi_k({\bf r}_1) \phi_l({\bf r}_2) d{\bf r}_1 d{\bf r}_2 .
\label{electron-hole-exchange}
\end{eqnarray}

\section{Exciton recombination}

In order to calculate the photoluminescence spectrum, one needs to calculate eigenstates of both $N$ exciton and $N-1$
exciton systems. At low temperature, only the ground state and a few excited states of the $N$ exciton system are
required. However, in order to obtain the spectrum over a broad energy range, a larger number of eigenstates of the
$N-1$ exciton system has to be calculated. In general, about $1000-2000$ eigenstates of the $N-1$ exciton system are
required to cover transitions occurring in the $s$ and $p$ shells.

Let us begin with recombination of non-interacting electrons. 
The momentum matrix element between an electron state $\phi_e = \sum_n
\psi^e_n u_n $ and a hole state $\phi_h = \sum_n \psi^h_n u_n$ is given by
\begin{equation}
\langle \phi_h | {\bf e}\cdot\hat{\bf p} | \phi_e \rangle = \sum_{mn} 
\langle u_n |{\bf e}\cdot\hat{\bf p}| u_m \rangle
\langle \psi^h_n | \psi^e_m \rangle + \sum_m \langle \psi^h_m |{\bf e}\cdot\hat{\bf p}| \psi^e_m \rangle,
\label{momentum-matrix-element}
\end{equation}
Here, $\psi_n$'s are the envelop functions and the basis functions $u_n$'s are chosen as eight uncoupled spin-orbitals,
i.e., $|s\uparrow\rangle$, $|x\uparrow\rangle$, $|y\uparrow\rangle$, $|z\uparrow\rangle$, $|s\downarrow\rangle$,
$|x\downarrow\rangle$, $|y\downarrow\rangle$, and $|z\downarrow\rangle$. If we neglect the contribution from the
envelope-function part of the wave function $\sum_m \langle \psi^h_m |{\bf e}\cdot{\bf p}| \psi^e_m \rangle$
\cite{KP:Sheng2001}, it can be further simplified as
\begin{eqnarray}
\langle \phi_h | \hat{p}_x | \phi_e \rangle &=& iP_0 \cdot \bigl[
\langle \psi^h_{x\uparrow} | \psi^e_{s\uparrow} \rangle + \langle \psi^h_{x\downarrow} | \psi^e_{s\downarrow} \rangle )
\cr 
&-& \langle \psi^h_{s\uparrow} | \psi^e_{x\uparrow} \rangle - \langle \psi^h_{s\downarrow} | \psi^e_{x\downarrow} 
\rangle \bigr] , \cr
\langle \phi_h | \hat{p}_y | \phi_e \rangle &=& iP_0 \cdot \bigl[
\langle \psi^h_{y\uparrow} | \psi^e_{s\uparrow} \rangle + \langle \psi^h_{y\downarrow} | \psi^e_{s\downarrow} \rangle 
\cr 
&-& \langle \psi^h_{s\uparrow} | \psi^e_{y\uparrow} \rangle - \langle \psi^h_{s\downarrow} | \psi^e_{y\downarrow} 
\rangle \bigr] ,
\label{momentum-matrix-element-simplified}
\end{eqnarray}
where $iP_0 = \langle s | \hat{p}_x | x \rangle = \langle s | \hat{p}_y | y \rangle$ denotes the coupling between the
conduction and valence bands. For circular polarization $\sigma^+$ or $\sigma^-$, the momentum matrix element is then
given by $p_{he}^\pm = \frac{1}{\sqrt{2}}( \langle \phi_h | \hat{p}_x | \phi_e \rangle \mp i \langle \phi_h | \hat{p}_y
| \phi_e \rangle )$.

In the single-band effective-mass method, the Bloch functions for the heavy hole are $ u_{h\uparrow} =
\frac{1}{\sqrt{2}} (|x\uparrow\rangle + i|y\uparrow\rangle) $ for $j_z=3/2$ and $ u_{h\downarrow} = \frac{1}{\sqrt{2}}
(|x\downarrow\rangle - i|y\downarrow\rangle) $ for $j_z=-3/2$. Hence, we have $ p_{he}^- = \langle \phi_{h\uparrow} |
\hat{p}^- | \phi_{e\uparrow} \rangle = \langle u_{h\uparrow} | \frac{1}{\sqrt{2}}( \hat{p}_x + i\hat{p}_y ) | s\uparrow
\rangle \langle \psi_h | \psi_e \rangle = -iP_0 \langle \psi_h | \psi_e \rangle $. It is straightforward to show that $
p_{he}^- = p_{he}^+ $.

The intensity of photoluminescence from the recombination of one electron-hole pair in a $N$-exciton state is defined
as \cite{MultiCI:Hawrylak1999,CI:Cheng}
\begin{eqnarray}
I_{\sigma_\pm}(h\nu) = && \sum_i f(E_N^i) \sum_f | \langle C_{N-1}^f | P^-_{\sigma_\pm} | C_N^i \rangle |^2 \cr
&& \cdot \delta( E_N^i - E_{N-1}^f - h\nu) ,
\label{recombination-spectrum}
\end{eqnarray}
where $C_N^i$ is the $i$-th eigenstate of the $N$-exciton system. Note that $\langle C_{N-1}^f | P^-_{\sigma_\pm} |
C_N^i \rangle$ coherently sums all the possible recombinations, therefore, the interference effect may play an 
important role.

The probability function is defined as $ f(E_N^i) = \exp(-E_N^i/\kappa T)/\sum_j \exp(-E_N^j/\kappa T)$. The operator
$P^-_{\sigma_\pm}$ describes all the possible electron-hole recombination, namely,
\begin{equation}
P^-_{\sigma_\pm} = \sum_{nm} p_{nm}^\pm h_n c_m ,
\label{recombination-operator}
\end{equation}
In the absence of magnetic field, we  have $I_{\sigma_+}(E) = I_{\sigma_-}(E)$.

\section{Results and Discussion}

We now illustrate our method by a calculation for a model structure of $\mbox{In}_{\mbox{\scriptsize
0.5}}\mbox{Ga}_{\mbox{\scriptsize 0.5}}\mbox{As/GaAs}$ disklike SAQD characteristic of SAQD's grown using In-flush
method\cite{Inflush:Wasilewski}. The dot has diameter 25.4 nm along the base and 2.3 nm height along the growth
direction. The composition and dimensions of the dot are chosen such that its emission spectrum is similar to the one
observed in the experiment of Raymond {\it et al} \cite{EXP:Raymond}. The strain distribution is calculated by the
continuum elasticity theory \cite{Strain:Pryor, Strain:Tadic} on a large cubic finite-difference mesh that has 120 nm
along each dimension and Dirichlet boundary condition on each side in order to ensure that the strain is fully relaxed.

\begin{figure}
\vspace{5mm}
  \includegraphics[width=3in]{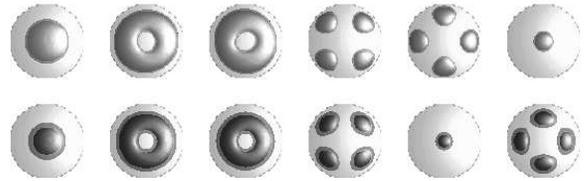}\\
  \caption{ Two-dimensional plot of the density of first six electron (first row) and hole (second row) wave functions 
calculated by the multiband $k\cdot p$ method. }
\label{single-particle-states}
\end{figure}

\begin{figure}
\vspace{5mm}
  \includegraphics[width=3in]{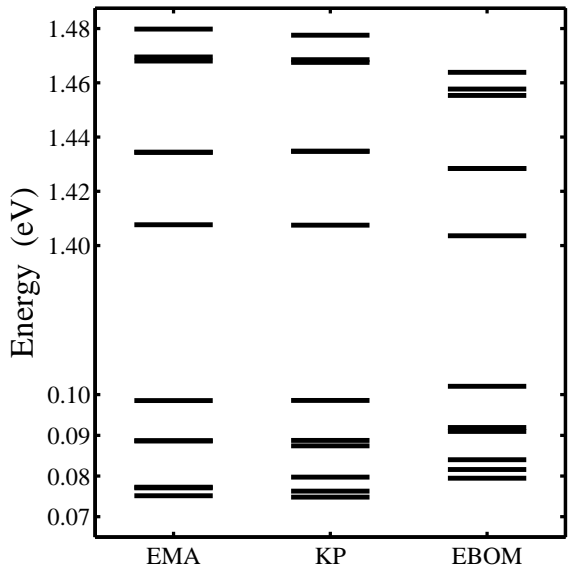}\\
  \caption{ Calculated energy levels of the $\mbox{In}_{\mbox{\scriptsize 0.5}}\mbox{Ga}_{\mbox{\scriptsize
0.5}}\mbox{As/GaAs}$ self-assembled quantum dot by the eight-band $k\cdot p$ (KP) method and EBOM. Also shown are the 
energy levels fitted for the effective-mass approximation (EMA). }
\label{single-particle-levels}
\end{figure}

In Fig. \ref{single-particle-states}, we show the probability density of the first six electron (in the upper row) and
six hole states (lower row) calculated by the multiband $k\cdot p$ method. The corresponding energy levels are plotted
in Fig. \ref{single-particle-levels}. The material parameters used in the calculation are taken from Ref.
\cite{KP:Pryor}.

The circular symmetry of the single-particle states is found basically preserved due to the small shear strain and weak
piezoelectric potential in this intermixed quantum dot. Hence, the states in the conduction band and valence bands are
seen to group into three shell, respectively. In the second shell, the shear strain induces a small splitting of
1.3~meV between the two $p$-like valence-band states. In the third shell, the splitting is about 3~meV and the disklike
geometry is responsible for the splitting between the two $3d$ states and the $2s$ state. It should also be noted that 
the ordering of the electron states in this shell is different from that of the hole states.

In the single-band effective mass calculation, the four effective mass parameters are chosen to fit the spectrum to
that obtained by the $k\cdot p$ method, giving ( in unit of free electron mass $m_0$ ) $m^e_\parallel = 0.060$,
$m^e_\perp = 0.070$, $m^h_\parallel = 0.27$, and $m^h_\perp = 0.30$. It is noted that the electron effective mass in
the dot is larger than the bulk value $0.045$ for $\mbox{In}_{\mbox{\scriptsize 0.5}}\mbox{Ga}_{\mbox{\scriptsize
0.5}}\mbox{As}$ and approaches the value in bulk GaAs. The anisotropy in the effective-mass tensor in the valence
bands, $0.30/0.27$, is greatly reduced comparing with the bulk value, $0.29/0.074$, i.e., the holes are much heavier in
the plane perpendicular to the growth direction. Similar findings that the effective mass of electrons in quantum dots
exceeds the value in the corresponding bulk dot material and approach that in the bulk matrix material and the in-plane
component of the effective mass of holes becomes much lighter have been reported \cite{EXP:Drexler, EXP:Schulhauser}.

In valence bands, the heavy hole and light hole are decoupled by the biaxial strain. For dots of small height, the
biaxial strain is almost constant inside the structure, hence, the low-lying states in valence bands are mostly
heavy-hole states. This is the reason why these states can be fitted by using single-band approximation. For thick
dots, the band edges of heavy hole and light hole may cross each other due to the fact that the biaxial strain changes
its sign inside the structure, which results in more light-hole components in the hole states in these dots.

The energy levels calculated by EBOM differs from those by the $k\cdot p$ method, especially for the high-lying states.
The shell separations by EBOM are smaller than those by the $k\cdot p$ method. For example, in the conduction band,
$E_{s-p}$ (separation between $s$ and $p$ shell) is 24.8 meV by EBOM and 27.3 meV by the $k\cdot p$ method,
respectively. The averaged separations $E_{p-d}$ between different models are even larger, 30.6 meV from EBOM and 36.4
meV from the $k\cdot p$ method.

As shown in Fig. \ref{bands-diagram}, the band structure predicted by the eight-band $k\cdot p$
method and EBOM match only in the region close to the $\Gamma$ point. Although the lateral dimensions of the
quantum-dot structure are large, it is very thin (2.3 nm) along the growth direction. Hence, the confined electron and
hole states include components with large $k_z$, which results in  different energies for these states.
For dots with larger height, there is very little discrepancy found between the two methods \cite{EBOM:Sun}.

\subsection{Polarization of single-particle states}

In the framework of the envelope function formalism, it is possible to separate the components of an envelope function
into two groups when the spin-orbit interaction is not very strong. One group consists of components for spin up
basis functions, $|s\uparrow\rangle$, $|x\uparrow\rangle$, $|y\uparrow\rangle$, and $|z\uparrow\rangle$, and the other
consists of components for spin down basis functions. We define polarization of state $\phi$ as
\begin{equation}
p = \int |\psi_{s\uparrow}({\bf r})|^2 + |\psi_{x\uparrow}({\bf r})|^2 +
         |\psi_{y\uparrow}({\bf r})|^2 + |\psi_{z\uparrow}({\bf r})|^2 ~d{\bf r}.
\label{parity}
\end{equation}
A state is polarized if either $p \approx 1$ (a `spin' up state) or $p \approx 0$ (a `spin' down state). Apparently,
there is little overlap between the polarized states with different polarization. 

A careful examination of the calculated single-particle states shows that all the single-particle states in the
conduction band consist of less than $1\%$ component from the split-off band while for valence band states it is less
than $5\%$. Hence, the mixture between spin up and spin down components in any of these states should be very small,
i.e., they are polarized.

However, in the absence of magnetic fields, all the single-particle states calculated by the $k\cdot p$ method and EBOM
are doubly degenerate due to the time-reversal symmetry \cite{KP:Stier}. Instead of having two degenerate states, the
numerical calculation can only give one state from a random linear combination of the two polarized and degenerate
states. Because of this degeneracy, most of the calculated single-particle states are found not polarized. By applying
a small magnetic field (1 mT) along the growth direction, this time-reversal symmetry can be removed and polarized
single-particle states are recovered. The eight-band $k\cdot p$ Hamiltonian is modified \cite{KP:Ram-Mohan} to include
the effects of magnetic fields. For EBOM, we introduce Peierls phase factor \cite{EBOM:Graf} to include the magnetic
field in the Hamiltonian.

\subsection{Electron-hole exchange interaction}

The ground state of a single exciton is a dark doublet separated from a bright doublet by the exchange energy. A dark
(bright) exciton state is dominated by a configuration of an electron and a hole in their respective ground state with
the opposite (same) spin. The bright doublet has a higher energy due to the electron-hole exchange energy. Because of
the relatively large size of SAQDs, the electron-hole exchange interaction causes a very small correction to exciton
states. It can be measured from the fine structure of single exciton recombination spectrum \cite{EXP:Bayer2002}.

An accurate calculation of the electron-hole exchange will require knowledge of both electron and hole states and the
dielectric function at a microscopic level \cite{MultiCI:Franceschetti, MultiCI:Bester}. However, for SAQDs, the
exchange energy can be estimated by using the multiband $k\cdot p$ theory or EBOM, where the electron-hole exchange
interaction arises from the mixing between the conduction and valence bands. The calculation for our structure shows
that the separation between the dark and bright doublets of a single exciton is $74.6~\mu\mbox{eV}$, which includes an
electron-hole exchange energy $63.2 ~ \mu \mbox{eV}$ and the correlation effect. It gives a fairly good agreement with
the value derived from the experiment \cite{EXP:Bayer2002} on similar samples, considering the approximation made in
the theory, and the uncertainty in the dot size, shape and composition in the experiment. Because of the small
contribution of the electron-hole exchange interaction, we neglect it in the subsequent calculation.

\subsection{Addition energies and hidden symmetry in multi-exciton complexes}

Due to the presence of quasi-shell structure in the single-particle energy spectrum, we chose the first 12 electron
and 12 hole states (with quasispins), which form the first three shells in conduction and valence bands, respectively,
to build the multi-exciton configurations.

In order to reduce the total number of configurations which grows factorially with the size of single-particle basis
set, we impose an additional constraint on the exciton configurations, i.e., the sum of the electron quasi-spins
should be equal to that of the hole quasispins \cite{CI:Cheng}. We also apply a truncation according to the
Hartree-Fock energies of configurations in order to limit the total number of the configurations to less than $50,000$.

Hidden symmetry \cite{EXP:Bayer2000, MultiCI:Wojs1996, MultiCI:Hawrylak1999, CI:Cheng} is a good approximation in a
multi-exciton system with degenerate single-particle states and symmetric electron-electron, electron-hole, and
hole-hole interactions. It predicts that the chemical potential i.e. the energy required to add an electron-hole pair
to the system, is independent of the number of excitons.

The symmetry between electron-electron, electron-hole, and hole-hole interactions is broken because the hole states are
usually more confined than the electrons in the conduction bands. Hence, the hole-hole interaction is generally
stronger than the electron-electron interaction. The calculation by the multiband $k\cdot p$ (single-band
effective-mass) method shows $V^{ee}_{1111} = 14.4 (14.9) ~ \mbox{meV}, ~ V^{hh}_{1111} = 16.7 (19.4) ~ \mbox{meV},
~\mbox{and}~ V^{he}_{1111} = 15.4 (16.6) ~ \mbox{meV}$. The EBOM gives similar values, which are 13.9~meV, 16.3~meV,
and 14.9~meV, respectively. In the case of either $k\cdot p$ or EBOM, the hole-hole interaction is stronger than the
electron-electron interaction by about 15\%. The hole states from the single-band calculation are found more confined
than those from the two multiband calculations.

\begin{figure}
\vspace{5mm}
  \includegraphics[width=3in]{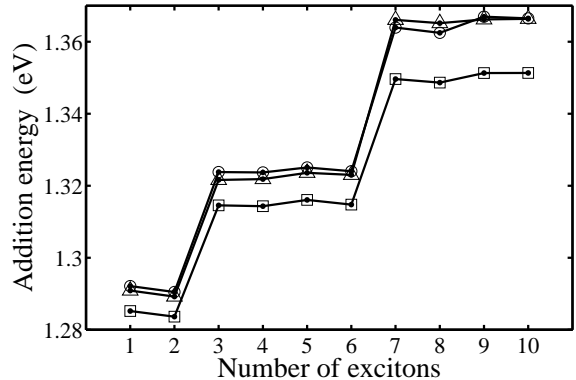}\\
  \caption{ Addition energy spectra $\mu(N) = E_g(N)-E_g(N-1)$, calculated by the single-band effective-mass
approximation (triangular dots), the multiband $k\cdot p$ method (circular dots) and EBOM (square dots) for the
$\mbox{In}_{\mbox{\scriptsize 0.5}}\mbox{Ga}_{\mbox{\scriptsize 0.5}}\mbox{As/GaAs}$ disklike quantum dot.}
\label{addition-spectrum}
\end{figure}

Fig. \ref{addition-spectrum} plots the calculated addition energies for different number of excitons. Both the $k\cdot
p$ method and EBOM give similar result except that the EBOM predicts lower values. The result of the single-band
effective-mass approximation is found very close to that by the $k\cdot p$ method. A clear plateau structure can be
found associated with the shell structure, which is an apparent signature of hidden symmetry. The fluctuation in the
addition energies within the same shell is not larger than that in the single-particle energies or the difference among
the electron (hole)-electron (hole) interactions. The $k\cdot p$ method gives the largest fluctuation 1.4 meV in the
$p$ shell and 4.5 meV in the $d$ shell, while EBOM gives $1.7$ meV and $2.7$ meV, respectively. It is therefore seen
that the hidden symmetry in our structure is well preserved, and not sensitive either to the splitting of $p$ and $d$
shells or the asymmetric interactions.

\subsection{Emission spectra}

\begin{figure}
\vspace{5mm}
  \includegraphics[width=3in]{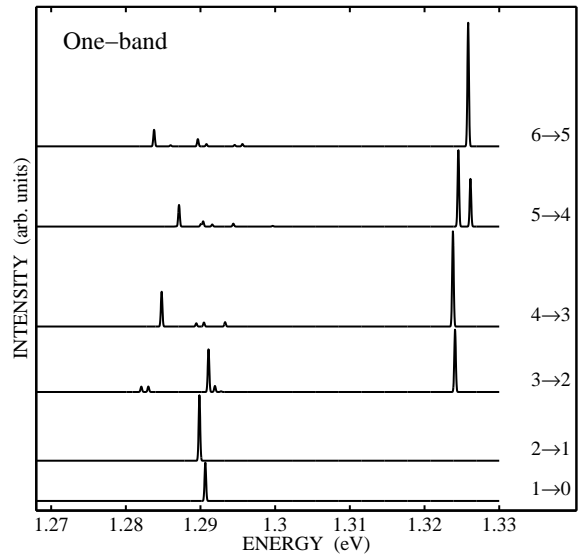}\\
  \caption{ Excitonic emission spectra calculated by the single-band effective-mass method. }
\label{singleband}
\end{figure}

\begin{figure}
\vspace{5mm}
  \includegraphics[width=3in]{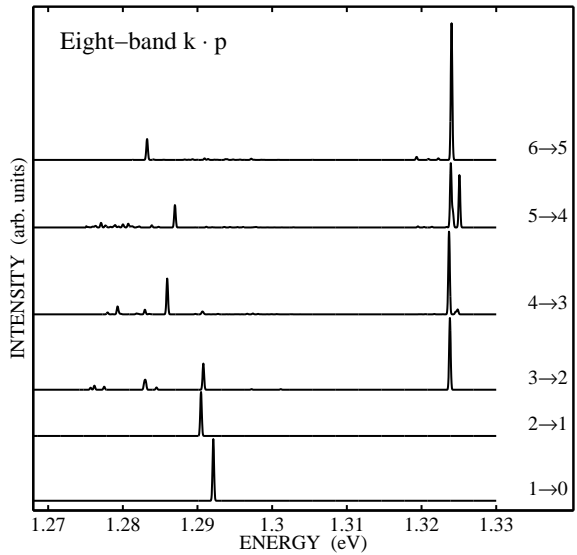}\\
  \caption{ Excitonic emission spectra calculated by the multiband $k\cdot p$ method. }
\label{kpbands}
\end{figure}

\begin{figure}
\vspace{5mm}
  \includegraphics[width=3in]{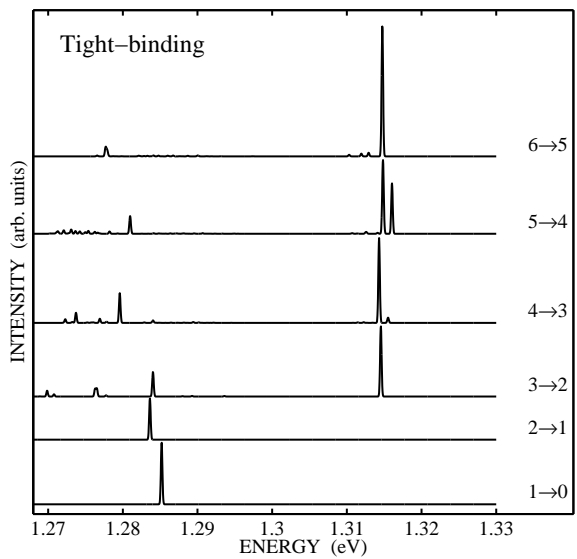}\\
  \caption{ Excitonic emission spectra calculated by EBOM. }
\label{ebbands}
\end{figure}

Figs. \ref{singleband}, \ref{kpbands}, and \ref{ebbands} show the emission spectra calculated by the single-band
effective-mass method, the eight-band $k\cdot p$ method, and EBOM, respectively, for up to six excitons. As the $k\cdot
p$ method and EBOM are shown to give very similar results except for overall shifts in transition energies, we will
confine our attention to analyzing the difference between the results by the single-band and the multiband $k\cdot p$
calculations. It should be mentioned that the spectra in Figs. \ref{singleband} and \ref{kpbands} are plotted in
different scale, only the relative intensity between the spectra in the same figure is relevant.

The difference between the single-band and multiband calculations concentrates in the $s$ shell as the emission peaks
in the $p$ shell are found similar between Figs. \ref{singleband} and \ref{kpbands}. Compared with the single-band
calculation, the band-mixing effect enables more configurations from the multiband single-particle states to be coupled
through the Coulomb interaction. It results in more possible yet weak transitions in the low energy end of the spectra,
as are found in Figs. \ref{kpbands} and \ref{ebbands} when the number of excitons is larger than 4.

\begin{figure}
\vspace{5mm}  
  \includegraphics[width=3.4in]{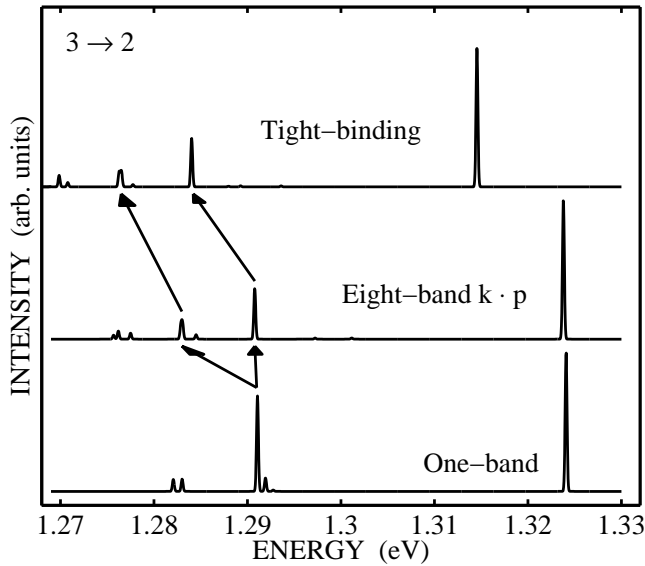}\\
  \caption{ Emission spectra from three-exciton complex calculated by different methods. }
\label{three-exciton}
\end{figure}

This is illustrated in Fig. \ref{three-exciton}. In the emission spectra from the three-exciton complex (3X), all the
three methods give one strong peak in the $p$-shell region and exhibit different structure in the $s$-shell region. The
single-band calculation gives one peak with high intensity and three other small peaks in the $s$-shell region while
both the $k\cdot p$ method and EBOM show five peaks with visible intensities.

The initial state of the recombination from 3X is its ground state and the final states associated with peaks found in
the $s$-shell region are excited biexcition states \cite{MultiCI:Hawrylak1999}. As there is little difference among the
3X ground states calculated by different methods, it is the excited biexciton states that account for the different
structure in the emission spectra.

In the single-band calculation, the emission peak with high intensity in the $s$-shell region is associated with three
biexciton states where both the two electrons and the two holes are in a triplet configuration (one in the $s$ shell
and the other in the $p$ shell). In these states, the total spin $S(=1)$ and its z-component $S_z$ of the two electrons
are the same as those of the two holes. As the spin-orbit interaction is absent in the single-band calculation, these
three biexciton states of different $S_z$ are degenerate and give rise to only one peak in the $s$-shell region. 

In the presence of spin-orbit interaction which is taken into account in both the $k\cdot p$ method and EBOM, the
degeneracy among these three biexciton states is partially lifted, i.e., the biexciton state of $|S_z|=1$ has a
different energy from that of $Sz=0$. It gives rise to two splitted peaks in the $s$-shell region, as illustrated in
Fig. \ref{three-exciton}. The three other smaller peaks are associated with those excited biexciton states in
singlet-singlet or singlet-triplet configurations, which are less affected by the spin-orbit interaction, and can be
seen in the all spectra.

\section{Conclusions}

In conclusion, we have presented a multiband microscopic theory of many-exciton complexes in self-assembled quantum
dots. Three methods: single-band effective-mass approximation, the multiband $k\cdot p$ method, and the
tight-binding-like EBOM, are used to obtain single-particle states. We expand the many-body wave functions of $N$
electrons and $N$ valence holes in the basis of Slater determinants. The Coulomb matrix elements are evaluated using
statically screened interaction and the correlated $N$-exciton states are obtained by the configuration interaction
method. We apply the theory to the excitonic emission spectrum in InAs/GaAs self-assembled quantum dots and
successfully compare the results of the single-band effective-mass approximation with those obtained by using the of
$k\cdot p$ and tight-binding methods.

\acknowledgments 
This work is supported by the NRC HPC project, NRC-HELMHOLTZ grant, and CIAR. The authors would like to thank M.
Korkusi$\acute{n}$ski, S. Patchkovskii and J. Tse for discussions.

\appendix

\section{Coulomb matrix elements}

In this appendix, we describe an efficient approach to calculate Coulomb matrix elements numerically. In the
configuration interaction method, the properties of the system are given by the single particle spectrum and by the
Coulomb matrix elements defined as two-electron integrals (see Eq. [\ref{two-electron-integral}]). One possible way to
calculate these six-dimensional integrals is to first solve Poisson equation, than calculate reduced three-dimensional
integrals \cite{CI:Heitz}. The calculation is repeated for each integrals. For the calculation involving 12 electron
states and 12 hole states, the total number of integrals is almost ten thousands. Here we propose an algorithm that 
does not require calculation for each integrals.

Within the envelope function formalism, the wave functions of single-particle states are expressed as a linear
combination of Bloch sums,
\begin{equation}
\phi_i({\bf r}) = \frac{1}{\sqrt{N}} \sum_n \sum_{\bf R} \psi_{in}({\bf R}) u_n({\bf r}-{\bf R}),
\label{envelopfunction}
\end{equation}
where $\psi_{in}$ is the $n$-th component of the envelop function $\psi_{i}$, $u_n({\bf r}-{\bf R})$ is the $n$-th
atomic orbital localized at unit cell ${\bf R}$ and $1/\sqrt{N} \sum_n u_n({\bf r}-{\bf R})$ is the corresponding
Bloch function. If we ignore the contribution from these localized atomic orbitals and replace ${\bf r}_1-{\bf r}_2$
with ${\bf R}_1-{\bf R}_2$ in Eq. \ref{two-electron-integral}, we have
\begin{eqnarray}
V_{ijkl} &=& \sum_{{\bf R}_1,{\bf R}_2} \sum_m \psi_{im}^\ast({\bf R}_1) \psi_{lm}({\bf R}_1) \cr
&& \cdot \frac{e^2}{4 \pi \epsilon\cdot|{\bf R}_1-{\bf R}_2|} \sum_n \psi_{jn}^\ast({\bf R}_2) 
\psi_{kn}({\bf R}_2) .
\label{two-electron-integral-in-calculation}
\end{eqnarray}  

We further transform a three-dimensional (3D) wave function $\psi({\bf R})$ into a column vector $\psi(r)$ by
mapping the 3D variable ${\bf R}$ onto a one-dimensional index $r$, a six-dimensional integral can be converted
into a vector-matrix multiplication,
\begin{equation}
V_{ijkl} = (\sum_m \psi_{im}^\ast \otimes \psi_{lm})^T \cdot U \cdot 
(\sum_n \psi_{jn}^\ast \otimes \psi_{kn}),
\label{integral2multi}
\end{equation}
where $\otimes$ is the direct multiplication (element by element) operator between two vectors. $U$ is the matrix with 
elements $U(r_1,r_2) = \frac{e^2}{4 \pi \epsilon\cdot|{\bf R}_1-{\bf R}_2|}$. The diagonal elements can be 
obtained by the integration of $1/R$ over a unit cell.

In order to use the optimized BLAS (Basic Linear Algebra Subprograms) library \cite{BLAS}, the above formulation can be
further vectorized as
\begin{equation}
J = \Phi^T \cdot U \cdot \Phi ,
\label{vectorization}
\end{equation}
where $\Phi_{r,\{ij\}} = \sum_n \psi_{in}^\ast(r) \psi_{jn}(r)$ is a matrix containing all the possible pairs of
two-particle wave functions. Due to the large dimension of matrix $U$, we make use of domain decomposition in the
numerical calculation to divide it into a number of smaller matrices and sum up the result of all the individual
multiplications.

\section{Companion configuration and additivity rule}

In this appendix, we point out how to use the additivity rule to construct configurations for multiexcitons. As a
multi-exciton complex contains two different particles, electrons and holes, the total number of possible
configurations is much larger than for electrons or holes separately. To circumvent this difficulty, we use the
following algorithm for construction of multi-exciton configurations.

The many-body Hamiltonian matrix constructed from the CI method is a sparse matrix. For a given configuration, there is
only a small number of configurations interacting with it, which are named as its {\it companion} configurations. Let
us first define the distance between configuration $C_i$ and $C_j$, $||C_i,C_j||$, as the total number of
single-particle states that the two configurations differ by. It is apparent that
\begin{equation}
\langle C_i | \hat{H} | C_j \rangle = 0,~\mbox{if}~||C_i,C_j||>2. 
\label{distance}
\end{equation}

An exciton configuration consists of a part for electron(s) and the other part for hole(s), i.e., $C_i^{ex} = \{C_i^e,
C_i^h\}$. The distance between two exciton configurations, $C_i^{ex}$ and $C_j^{ex}$, can be easily calculated by the
{\it additivity} rule, namely,
\begin{eqnarray}
D_{ij}^{ex} &=& ||C_i^{ex},C_j^{ex}|| = ||C_i^{e},C_j^{e}|| + ||C_i^{h},C_j^{h}|| \cr
&=& D_{ij}^{e} + D_{ij}^{h}.
\label{additivity-rule}
\end{eqnarray}

One can calculate the distance matrix $D_e$ and $D_h$ for the electron and hole configurations, respectively, and then
obtain the matrix $D_{ex}$ by using the additivity rule. When the number of single particle states (either electrons or
holes) is large, a cutoff is necessary to be applied to the total number of electron or hole configurations. Depending
on the memory available for the computation, it is set to be between 5000 and 10000. Once the distance matrix $D_{ex}$
for the exciton configurations is calculated, it is straightforward to construct the configuration interaction matrix
as the positions of all the none-zero matrix elements are known.

\end{document}